\documentclass[a4paper,11pt]{article}
\usepackage{pos}
\usepackage{setspace}
\usepackage{wrapfig}

\usepackage{cleveref}
\allowdisplaybreaks

\def\pdotq{p \cdot q}
\def\sdotq{s \cdot q}

\newcommand{\bv}[1]{{\mathbf{#1}}}
\newcommand{\bs}[1]{{\boldsymbol{#1}}}

\def\latmom{\left( \frac{2\pi}{L} \right)}
\def\llangle{\left\langle}
\def\rrangle{\right\rangle}
\newcommand{\ea}[1]{{\llangle #1 \rrangle}}

\title{The Compton amplitude and nucleon structure functions}

\manuallySeparateAuthors

\author*[a]{K.~Utku~Can}
\author{ for the QCDSF/UKQCD Collaboration}

\affiliation[a]{CSSM, Department of Physics, The University of Adelaide,
Adelaide SA 5005, Australia.}

\emailAdd{kadirutku.can@adelaide.edu.au}

\abstract{Structure functions are the essential objects for understanding the deep inelastic scattering processes, providing valuable insight into the partonic structure of hadrons. A direct calculation of the Compton amplitude provides a complementary way to accessing the structure functions, circumventing the operator mixing and renormalisation issues of the standard operator product expansion approach.
We describe the connection between the Compton amplitude and the structure functions, and describe a Feynman-Hellmann approach to calculate the amplitude directly. As an application, we extract the moments of transverse and longitudinal proton structure functions and study the power corrections. 
}

\FullConference{%
  The 39th International Symposium on Lattice Field Theory (Lattice2022),\\
  8-13 August, 2022 \\
  Bonn, Germany 
}

\begin{document}
\maketitle

\section{Introduction}
Calculating the structure functions from first principals poses several challenges for lattice QCD practitioners. Traditionally, lattice calculations make use of the operator product (OPE) expansion. However, it is known that in the OPE approach, contributions of leading-twist operators are inseparably connected with the contributions from operators of higher twist, due to operator mixing and renormalisation~\cite{Martinelli:1996pk}. In recent years, the focus has been on light-cone parton distribution functions (PDFs) accessible from the quasi-PDF approach introduced by Ji~\cite{Ji:2013dva}, which enables a direct investigation of the $x$-dependence of parton distributions. A detailed account of the quasi-PDF and related approaches is given in recent reviews~\cite{Lin:2017snn,Cichy:2018mum} and presented in plenary talks at the lattice conferences~\cite{Constantinou:2020pek,Cichy:2021lih}, highlighting the immense efforts and the progress of the lattice community. The operator mixing issue, however, remains a concern~\cite{Martinelli:1996pk,Rossi:2017muf,Braun:2018brg}. The majority of the investigations are limited to the leading-twist contributions, with fewer works on twist-3 contributions~\cite{Gockeler:2005vw,Bhattacharya:2020cen,Bhattacharya:2021moj}.    

In this contribution we describe an alternative and complementary approach that is being pursued by the QCDSF/UKQCD Collaboration, which is to directly calculate the forward Compton amplitude on the lattice in the space-like region. While the Compton amplitude is a 4-point correlation function, via an application of the Feynman-Hellmann approach we reduce this problem to a more straightforward analysis of 2-point functions. By working with the physical amplitude, the operator mixing and renormalization issues, and the restriction to light-cone operators are circumvented. Given the Compton amplitude is known sufficiently accurately, we can expect to estimate the power corrections in structure functions, i.e. quantify the target mass corrections and estimate the contributions from higher-twist operators, which could be useful for global PDF analyses. In principle, the $x$-dependence of the structure functions can be recovered~\cite{PhysRevLett.118.242001}, although in practice this requires tackling an inverse-problem~\cite{Horsley:2020ltc}. Although our focus is the Compton amplitude in forward kinematics, this approach is applicable to off-forward kinematics enabling an investigation of the generalised parton distributions~\cite{Alec:2021lkf,Hannaford-Gunn:2022lez,Alec:2022lat}.

We give an overview of the relation between the Compton amplitude and the moments of structure functions in \Cref{sec:comp}, followed by a summary of the application of the Feynman-Hellmann theorem in \Cref{sec:fh}. The technical details follow in \Cref{sec:en,sec:g12}. We present some selected results on the moments of nucleon structure functions and discuss the power corrections in \Cref{sec:res}.   

\section{The Compton tensor and the moments of structure functions} \label{sec:comp}
The starting point is the forward Compton amplitude described by the time ordered product of electromagnetic currents sandwiched between nucleon states, 
\begin{align} 
    \label{eq:compamp}
    T_{\mu\nu}(p,q) =& \int d^4z\, e^{i q \cdot z} \rho_{s s^\prime} \ea{p,s^\prime \left| 
    \mathcal{T}\left\{ \mathcal{J}_\mu(z) \mathcal{J}_\nu(0) \right\} \right|p,s},
\end{align}
where $p$ is the momentum and $s$ is the spin of the nucleon, $q$ is the momentum of the virtual photon, and $\rho$ is the polarisation density matrix. For parity-conserving processes that involve conserved currents the Compton tensor is parametrised in terms of four Lorentz-invariant scalar functions, $\mathcal{F}_1$, $\mathcal{F}_2$, $\mathcal{G}_1$ and $\mathcal{G}_2$ as follows
\begin{align} 
    T_{\mu\nu}(p,q) &= T_{\{\mu\nu\}}(p,q) + T_{[\mu\nu]}(p,q)\\
    \label{eq:compamp_tensor}
    T_{\{\mu\nu\}}(p,q) &= \left( -g_{\mu\nu} + \frac{q_\mu q_\nu}{q^2} \right) \mathcal{F}_1(\omega,Q^2) + \left( p_\mu - \frac{p \cdot q}{q^2}q_\mu \right) \left( p_\nu - \frac{p \cdot q}{q^2}q_\nu \right) \frac{\mathcal{F}_2(\omega,Q^2)}{p \cdot q} \\
    \label{eq:compamp_tensor_pol}
    T_{[\mu\nu]}(p,q) &= i \varepsilon^{\mu\nu\alpha\beta} \frac{q_\alpha}{\pdotq} 
    \left[ \mathcal{G}_1(\omega,Q^2) s_\beta + \mathcal{G}_2(\omega,Q^2) \left(s_\beta - \frac{\sdotq}{\pdotq} p_\beta \right) \right],
\end{align}
where we have separated the symmetric and antisymmetric parts, and $Q^2 = -q^2$ and $\omega = 2 (\pdotq)/Q^2$. Here, $\varepsilon^{0123}=1$, and $s_\mu$ is the spin vector of the polarised target satisfying $s^2 = -M^2$ ($M$ being the mass of the nucleon) and $s \cdot p = 0$. 

The Compton structure functions are related to the corresponding ordinary structure functions via the optical theorem, which states
\begin{align}
    \operatorname{Im}\mathcal{F}_{1,2}(\omega,Q^2) = 2\pi F_{1,2}(x,Q^2), \\ 
    \operatorname{Im}\mathcal{G}_{1,2}(\omega,Q^2) = 2\pi g_{1,2}(x,Q^2).
\end{align}
Making use of analyticity, crossing symmetry and the optical theorem, we can write dispersion relations for $\mathcal{F}$ and $\mathcal{G}$ and connect them to the inelastic structure functions,
\begin{align}
    \label{eq:compomega12} 
    \overline{\mathcal{F}}_1(\omega,Q^2)= 2\omega^2 \int_0^1 dx \frac{2x \, F_1(x,Q^2)}{1-x^2\omega^2-i\epsilon}, \quad
    \mathcal{F}_2(\omega,Q^2)= 4\omega \int_{0}^1 dx\, \frac{F_2(x,Q^2)}{1-x^2\omega^2-i\epsilon}, \\
    \label{eq:compomega12_pol}
    \mathcal{G}_1(\omega,Q^2)= 4\omega \int_0^1 dx \frac{g_1(x,Q^2)}{1-x^2\omega^2-i\epsilon}, \quad
    \mathcal{G}_2(\omega,Q^2)= 4\omega \int_{0}^1 dx\, \frac{g_2(x,Q^2)}{1-x^2\omega^2-i\epsilon},
\end{align}
where we will use $\overline{\mathcal{F}}_i(\omega,Q^2) = \mathcal{F}_i(\omega,Q^2)-\mathcal{F}_i(0,Q^2)$ throughout to denote a once subtracted function. Additionally, a once-subtracted dispersion relation for the longitudinal structure function $F_L(x)$ is written as,
\begin{equation}\label{eq:compomegaL}
    \overline{\mathcal{F}}_L(\omega,Q^2) \equiv \mathcal{F}_L(\omega,Q^2) + \mathcal{F}_1(0,Q^2) = \frac{8M_N^2}{Q^2} \int_0^1 dx F_2(x,Q^2) + 2\omega^2 \int_0^1 dx \frac{F_L(x,Q^2)}{1 - x^2 \omega^2 -i\epsilon},
\end{equation}
where,
\begin{align}
    \label{eq:FL_x}
    F_L(x,Q^2) &= \left( 1 + \frac{4 M_N^2}{Q^2} x^2 \right) F_2(x,Q^2) - 2xF_1(x,Q^2),
\end{align}
with $M_N$ the mass of the nucleon. Note that a subtraction is necessary, given the high-energy behaviour of $F_1$. Although we are only concerned with subtracting it away, understanding the subtraction function is an interesting subject in itself. Related discussions on the subtraction function can be found in~\cite{Walker-Loud:2012ift,Hagelstein:2020awq,Lozano:2020qcg,HannafordSankey:2021lat}. As $Q^2 \to \infty$, \Cref{eq:FL_x} reduces to the familiar Callan-Gross relation, $F_L(x) \to F_2(x) - 2xF_1(x)$, which vanishes in the quark-parton model. 

Expanding the integrands in \Cref{eq:compomega12,eq:compomega12_pol,eq:compomegaL} at fixed $Q^2$ as a geometric series, the Compton structure functions can be expressed as infinite sums over the Mellin moments of the inelastic structure functions,
\begin{align} 
    \label{eq:ope_moments1}
    \overline{\mathcal{F}}_1(\omega,Q^2)&=\sum_{n=1}^\infty 2\omega^{2n} M^{(1)}_{2n}(Q^2), &&\text{with} \; M^{(1)}_{2n}(Q^2)= 2\int_0^1 dx\, x^{2n-1} F_1(x,Q^2), \\
    \label{eq:ope_moments2}
    \mathcal{F}_2(\omega,Q^2)&= \sum_{n=1}^\infty 4\omega^{2n-1} M^{(2)}_{2n}(Q^2), &&\text{with} \; M^{(2)}_{2n}(Q^2)= \int_{0}^1 dx\,x^{2n-2} F_2(x,Q^2), \\
    \label{eq:ope_momentsL}
    \overline{\mathcal{F}}_L(\omega,Q^2) &= \sum_{n=1}^\infty 2\omega^{2n}M^{(L)}_{2n}(Q^2), &&\text{with} \; M^{(L)}_{2n}(Q^2)= \int_{0}^1 dx\,x^{2n-2} F_L(x,Q^2), \\
    \label{eq:ope_moments1_pol}
    \mathcal{G}_{1,2}(\omega,Q^2)&= \sum_{n=1}^\infty 4\omega^{2n-1} \tilde{M}^{(1,2)}_{2n}(Q^2), &&\text{with} \; \tilde{M}^{(1,2)}_{2n}(Q^2)= \int_{0}^1 dx\,x^{2n-2} g_{1,2}(x,Q^2).
\end{align}
We note that the physical moments $M_{2n}$ that appear in \Cref{eq:ope_moments1,eq:ope_moments2,eq:ope_momentsL,eq:ope_moments1_pol} are dominated by their leading-twist contributions, i.e. the moments of PDFs, at asymptotically large $Q^2$. 

Finally, the unpolarised Compton structure functions $\mathcal{F}_1$ and $\mathcal{F}_2$ are accessed from the symmetric part of the Compton tensor via,
\begin{align}
    \label{eq:compF1}
    \mathcal{F}_1(\omega, Q^2) &= T_{\{33\}}(p,q), &&\text{for} \, \mu=\nu=3 \, \text{and} \, p_3=q_3=0, \\
    \label{eq:compF2}
    \frac{\mathcal{F}_2(\omega,Q^2)}{\omega} &= \frac{Q^2}{2 E_N^2} \left[ T_{\{00\}}(p,q) + T_{\{33\}}(p,q) \right],
    && \text{for} \, \mu=\nu=0 \, \text{and} \, p_3=q_3=q_0=0,
\end{align}
and the longitudinal Compton structure function $\mathcal{F}_L$ is constructed as,
\begin{equation} \label{eq:FL_comp}
    \mathcal{F}_L(\omega, Q^2) = -\mathcal{F}_1(\omega,Q^2) + \frac{\omega}{2} \mathcal{F}_2(\omega,Q^2) + \frac{2 M_N^2}{Q^2} \frac{\mathcal{F}_2(\omega,Q^2)}{\omega}.
\end{equation}
We discuss how to access $\mathcal{G}_1$ and $\mathcal{G}_2$ in \Cref{sec:g12}.

\section{The Feynman-Hellmann technique} \label{sec:fh}
An analysis of the Compton amplitude requires the evaluation of lattice 4-point correlation functions. However, this is not an easy task given the rapid deterioration of the signal for large time separations and the contamination due to excited states. The application of the Feynman-Hellmann theorem reduces the problem to a simpler analysis of 2-point correlation functions using the established techniques of spectroscopy. Our implementation of the second order Feynman-Hellmann method is presented in detail in~\cite{PhysRevD.102.114505}. Here, we outline its main aspects. 

We modify the fermion action with the following perturbing term,
\begin{equation}\label{eq:fh_perturb}
    S(\lambda) = S + \lambda \int d^3z \operatorname{cos}(\bv{q} \cdot \bv{z}) \, \mathcal{J}_{\mu}(z),
\end{equation}
where $\lambda$ is the strength of the coupling between the quarks and the external field, $\mathcal{J}_{\mu}(z) = Z_V \bar{q}(z) \gamma_\mu q(z)$ is the renormalised electromagnetic current coupling to the quarks along the $\mu$ direction, $\bv{q}$ is the external momentum inserted by the current and $Z_V$ is the renormalization constant for the local electromagnetic current. The perturbation is introduced on the valence quarks only, hence only quark-line connected contributions are taken into account in this work. For the perturbation of valence and sea quarks see~\cite{Chambers2015}.  

The main strategy to derive the relation between the energy shift and the matrix element is to work out the second-order derivatives of the two-point correlation function with respect to the external field from two complementary perspectives. A two-point correlation function projected to definite momentum in the presence of an external field is defined as,
\begin{equation} \label{eq:2pt}
    G_{\lambda}^{(2)}(\bv{p};t;\bs{\Gamma}) \equiv \int d^3 x e^{-i \bv{p} \cdot \bv{x}} \bs{\Gamma} \langle \Omega_\lambda | \chi(\bv{x},t) \bar\chi(0) | \Omega_\lambda \rangle,
\end{equation}
where $\bv{\Gamma}$ is the spin-parity projection matrix and $| \Omega_\lambda \rangle$ is the vacuum in the presence of the external field. The asymptotic behaviour of the correlator at large Euclidean time takes the familiar form,
\begin{equation} \label{eq:G2spec}
      G^{(2)}_{\lambda}(\bv{p};t;\bs{\Gamma}) \simeq A_\lambda(\bv{p}) e^{ - E_{N_{\lambda}}(\bv{p}) \, t },
\end{equation}
where $E_{N_{\lambda}}(\bv{p})$ is the energy of the ground state nucleon in the external field and $A_\lambda(\bv{p})$ the corresponding overlap factor.
Differentiating the perturbed nucleon correlator (\Cref{eq:G2spec}) with respect to energy, one finds a distinct temporal signature for the second-order energy shift, 
\begin{align}\label{eq:G2spec_deriv}
        \frac{\partial^2 G^{(2)}_\lambda(\bv{p};t)}{\partial \lambda^2} \bigg |_{\lambda=0} =& \left( \frac{\partial^2 A_\lambda(\bv{p})}{\partial \lambda^2}  - t A(\bv{p}) \frac{\partial^2 E_{N_\lambda}(\bv{p})}{\partial \lambda^2}\right) 
        e^{-E_{N}(\bv{p}) t},
\end{align}
where we have assumed that first-order perturbations of the energy vanish, as ensured by avoiding Breit-frame kinematics. The derivatives of $A_\lambda(\bv{p})$ and $E_{N_\lambda}(\bv{p})$ are assumed to be evaluated at $\lambda=0$. The first term corresponds to the shift in the overlap factor and the second order energy shift is identified in the $t$-enhanced (or time-enhanced) term.

By differentiating \Cref{eq:2pt} twice with respect to $\lambda$ in the path integral formalism and evaluated at $\lambda\to 0$, we find
\begin{align} \label{eq:G2PI_deriv}
    \left.\frac{\partial^2 G^{(2)}_\lambda(\bv{p};y)}{\partial \lambda^2} \right |_{\lambda=0} = 
    &\int d^3x\, e^{-i \bv{p} \cdot \bv{x}}\bv{\Gamma}  
    \left[\ea{\chi(\bv{x},t) \overline{\chi}(0) \left(\frac{\partial S(\lambda)}{\partial \lambda}\right)^2} 
    + \cdots \right],
\end{align}
where ellipsis denote the terms that do not lead to a time-enhanced term. Inserting the explicit form of the external electromagnetic current (\Cref{eq:fh_perturb}) and following the algebra, we arrive at,
\begin{align} \label{eq:fh_pifin}
    &\left. \frac{\partial^2 G^{(2)}_\lambda(\bv{p};t)}{\partial \lambda^2} \right |_{\lambda=0} \hspace{-2.4mm} = 
    t A(\bv{p}) \frac{e^{-E_{N}(\bv{p}) t}}{2 E_{N}(\bv{p})}
    \llangle N(\bv{p}) 
    \left| \int d^4 z \left(e^{iq\cdot z}+e^{-iq\cdot z}\right) \mathcal{J}_{\mu}(z) \mathcal{J}_{\mu}(0) \right| N(\bv{p}) \rrangle + \ldots,
\end{align}
where the subleading terms are suppressed by the ellipsis.

Finally, matching the time-enhanced terms of this form with \Cref{eq:G2spec_deriv}, one arrives at the desired relation between the energy shift and the matrix element describing the Compton amplitude,
\begin{equation} \label{eq:secondorder_fh}
    \left. \frac{\partial^2 E_{N_\lambda}(\bv{p})}{\partial \lambda^2} \right|_{\lambda=0} = - \frac{T_{\mu\mu}(p,q) + T_{\mu\mu}(p,-q)}{2 E_{N}(\bv{p})},
\end{equation}
where $T$ is the Compton amplitude defined in \Cref{eq:compamp}. \Cref{eq:secondorder_fh} is the principal relation that we use to access the flavour-diagonal, i.e. $uu$ and $dd$, pieces of the symmetric part (\Cref{eq:compamp_tensor}) of the Compton amplitude.

The above derivation can be generalised to mixed currents by including additional perturbing terms in \Cref{eq:fh_perturb} to study the flavour-mixed, i.e. $ud$, piece, and the antisymmetric part (\Cref{eq:compamp_tensor_pol}) of the Compton tensor. On that account, we make the modifications,
\begin{align}
    \label{eq:fh_perturb_mixed}
    S(\lambda) &= S + \lambda_1 \int d^3z \operatorname{cos}(\bv{q} \cdot \bv{z}) \, \mathcal{J}_{\mu}(z)
    + \lambda_2 \int d^3y \operatorname{cos}(\bv{q} \cdot \bv{y}) \, \mathcal{J}_{\mu}(y), \\
    \label{eq:fh_perturb_pol}
    S(\lambda) &= S + \lambda_1 \int d^3z \operatorname{cos}(\bv{q} \cdot \bv{z}) \, \mathcal{J}_{\mu}(z)
    + \lambda_2 \int d^3y \operatorname{sin}(\bv{q} \cdot \bv{y}) \, \mathcal{J}_{\nu}(y),
\end{align}
to access the flavour-mixed and flavour-diagonal antisymmetric pieces of the amplitude, respectively. Consequently, expressions analogous to \Cref{eq:secondorder_fh} are,
\begin{align}
    \label{eq:secondorder_fh_mixed}
    \left. \frac{\partial^2 E_{N_\lambda}(\bv{p})}{\partial \lambda_1 \partial \lambda_2} \right|_{\bs{\lambda}=0} = - \frac{T_{\mu\mu}(p,q) + T_{\mu\mu}(p,-q)}{2 E_{N}(\bv{p})}, \\
    \label{eq:secondorder_fh_pol}
    \left. \frac{\partial^2 E_{N_\lambda}(\bv{p})}{\partial \lambda_1 \partial \lambda_2} \right|_{\bs{\lambda}=0} 
    = \frac{T_{\mu\nu}(p,q) - T_{\mu\nu}(p,-q)}{2 E_{N}(\bv{p})},
\end{align}
corresponding to the modifications in \Cref{eq:fh_perturb_mixed,eq:fh_perturb_pol}, respectively, where the crossing relations $T_{\mu\mu}(p,q) = T_{\mu\mu}(p,-q)$ and $T_{\mu\nu}(p,q) = -T_{\mu\nu}(p,-q)$ ensure that we have non-vanishing Feynman-Hellmann relations.

\section{Extracting the energy shifts} \label{sec:en}
We can expand the perturbed energy in the limit $\lambda \to 0$,
\begin{align}
     E_{N_{\lambda}}(\bv{p}) &= E_N(\bv{p}) 
     + \lambda \left. \frac{\partial E_{N_{\lambda}}(\bv{p})}{\partial \lambda} \right|_{\lambda=0}
     + \frac{\lambda^2}{2 !} \left. \frac{\partial^2 E_{N_{\lambda}}(\bv{p})}{\partial \lambda^2} \right|_{\lambda=0}
     + \mathcal{O}(\lambda^3) \\
     &= E_N(\bv{p}) 
     + \Delta E^e_{N_{\lambda}}(\bv{p}) 
     + \Delta E^o_{N_{\lambda}}(\bv{p}),
\end{align} 
considering a single current insertion (e.g. \Cref{eq:fh_perturb}), where we have collected the terms even ($e$) and odd ($o$) in $\lambda$ to all orders in writing the second line. A similar Taylor expansion of the perturbed energy can be written for the double current insertion (e.g. \Cref{eq:fh_perturb_mixed}),
\begin{equation}
    E_{N_{\lambda}}(\bv{p}) = E_N(\bv{p}) 
    + \Delta E^{eo}_{N_{\lambda}}(\bv{p}) 
    + \Delta E^{oe}_{N_{\lambda}}(\bv{p}) 
    + \Delta E^{ee}_{N_{\lambda}}(\bv{p}) 
    + \Delta E^{oo}_{N_{\lambda}}(\bv{p}),
\end{equation}
where the term of interest is,
\begin{equation} \label{eq:enshift_oo}
    \Delta E^{oo}_{N_{\lambda}}(\bv{p}) = \lambda_1 \lambda_2 \left. \frac{\partial^2 E_{N_{\lambda}}(\bv{p})}{\partial \lambda_1 \partial \lambda_2} \right|_{\lambda=0} + \mathcal{O}(\lambda_1 \lambda_2^3) + \mathcal{O}(\lambda_1^3 \lambda_2),
\end{equation}
with respect to \Cref{eq:secondorder_fh_mixed,eq:secondorder_fh_pol}.

We construct the ratios,
\begin{align}
    \label{eq:ratio_fd}
    \mathcal{R}^{e}_{\lambda}(\bv{p};t;\Gamma_4) &\equiv \frac{G^{(2)}_{+\lambda}(\bv{p};t;\Gamma_4) G^{(2)}_{-\lambda}(\bv{p};t;\Gamma_4)}{\left( G^{(2)}(\bv{p};t;\Gamma_4) \right)^2} 
    \xrightarrow{t \gg 0} A_{\lambda}(\bv{p}) e^{-2\Delta E^e_{N_\lambda}(\bv{p}) \, t}, \\
    \label{eq:ratio_fm}
    \mathcal{R}^{oo}_{\lambda}(\bv{p};t;\bs{\Gamma}) &\equiv \frac{G^{(2)}_{(+\lambda,+\lambda)}(\bv{p};t;\bs{\Gamma}) G^{(2)}_{(-\lambda,-\lambda)}(\bv{p};t;\bs{\Gamma})}{G^{(2)}_{(+\lambda,-\lambda)}(\bv{p};t;\bs{\Gamma}) G^{(2)}_{(-\lambda,+\lambda)}(\bv{p};t;\bs{\Gamma})} 
    \xrightarrow{t \gg 0} A_\lambda(\bv{p}) e^{-4\Delta E^{oo}_{N_\lambda}(\bv{p}) \, t},
\end{align} 
in order to extract the second-order energy shifts, where $A_\lambda(\bv{p})$ is the overlap factor, which is irrelevant for the rest of the discussion. These ratios isolate the energy shifts only at even orders of $\lambda$.  Here, $G^{(2)}_{\lambda}$ and $G^{(2)}_{(\lambda_1, \lambda_2)}$ are the perturbed two-point functions with $|\lambda_1| = |\lambda_2| = |\lambda|$, and $G^{(2)}$ is the unperturbed one. The spin-parity projection matrices are defined as $\Gamma_4 \equiv (1 + \gamma_4)/2$ for an unpolarised positive-parity nucleon, and $\Gamma_+^{\hat{e}} \equiv \Gamma_4 (1 + \bs{\hat{e}} \cdot \bs{\gamma} \gamma_5)/2$ for a spin-up positive-parity nucleon polarised along the $\hat{e}$ direction. \Cref{eq:ratio_fd} is used for the flavour-diagonal pieces of the symmetric part of the Compton amplitude, i.e. $\mathcal{F}^{(uu,dd)}_{(1,2)}$, while \Cref{eq:ratio_fm} with $\bs{\Gamma} = \Gamma_4$ is used for the flavour-mixed piece, i.e. $\mathcal{F}^{ud}_{(1,2)}$.

\section{Separating the $\mathcal{G}_1$ and $\mathcal{G}_2$} \label{sec:g12}
Accessing the polarised Compton structure functions $\mathcal{G}_1$ and $\mathcal{G}_2$ is more involved. By choosing the currents along the $\mu=1$, and $\nu=2$ directions, and adopting the kinematics $q_1 = p_1 = 0$, we cancel the symmetric part (\Cref{eq:compamp_tensor}) and isolate the spin dependent part of the tensor,
\begin{equation}\label{eq:spindep_comp}
    T_{[12]}(p,q) = i \varepsilon^{1230} \frac{q_3}{\pdotq} 
    \left[ \mathcal{G}_1(\omega,Q^2) s_0 + \mathcal{G}_2(\omega,Q^2) \left(s_0 - \frac{\sdotq}{\pdotq} p_0 \right) \right],
\end{equation}
where we are left with the choices $\alpha=3$ and $\beta=0$ to have a non-vanishing amplitude since $q_0=0$ by construct in the Feynman-Hellmann approach. We rearrange \Cref{eq:spindep_comp} by using the four-spin vector in a boosted frame,
\begin{equation}
    s^\mu(\bv{p}) \equiv \left( \frac{\bv{\bs{\hat{e}}} \cdot \bv{p}}{M_N}, 
    \bv{\bs{\hat{e}}} + \frac{\bv{\bs{\hat{e}}} \cdot \bv{p}}{M_N (E_N(\bv{p}) + M_N)} \bv{p} \right),
\end{equation}
where $\bv{\bs{\hat{e}}} = \sigma M_N \hat{n}$, with $\hat{n}$ the quantisation axis and $\sigma = + 1$ for spin up, along with the substitutions $\varepsilon^{0123} = 1$, $p_0 = E_N(\bv{p})$, $a \cdot q \equiv a^\mu q_\mu= - \bv{a} \cdot \bv{q}$, and $\omega = 2(\bv{p} \cdot \bv{q})/Q^2$, into a more compact form,
\begin{equation} \label{eq:Te}
    T^{\hat{e}}_{[12]}(p,q) = C_1(\bv{\bs{\hat{e}}},\bv{p},\bv{q}) \frac{\mathcal{G}_1(\omega,Q^2)}{\omega} + C_2(\bv{\bs{\hat{e}}},\bv{p},\bv{q}) \frac{\mathcal{G}_2(\omega,Q^2)}{\omega},
\end{equation} 
with the coefficients,
\begin{align}
    C_1(\bv{\bs{\hat{e}}},\bv{p},\bv{q}) &= i \frac{2 q_3}{Q^2} \frac{\bv{\bs{\hat{e}}}.\bv{p}}{M_N}, \\
    C_2(\bv{\bs{\hat{e}}},\bv{p},\bv{q}) &= i \frac{2 q_3}{Q^2} \left[ \frac{\bv{\bs{\hat{e}}}.\bv{p}}{M_N} - E_N(\bv{p}) \left(\frac{\bv{\bs{\hat{e}}}.\bv{p}}{M_N (E_N(\bv{p})+M_N)} + \frac{\bv{\bs{\hat{e}}} \cdot \bv{q}}{\bv{p} \cdot \bv{q}} \right) \right].
\end{align}
Note that we have introduced a superscript $\hat{e}$ in \Cref{eq:Te} to keep track of the spin polarisations. Once we extract the energy shifts, $\Delta E_{N_{\lambda}}^{oo}(\bv{p})$, using \Cref{eq:ratio_fm} with $\bs{\Gamma} = \Gamma_+^{\hat{e}}$ and $\hat{e}$ along the $\hat{y}$ and $\hat{z}$ directions, and determine the Compton amplitude using \Cref{eq:secondorder_fh_pol}, we separate $\mathcal{G}_{1}$ and $\mathcal{G}_{2}$ by solving the system of linear equations,
\begin{equation}
    \bv{T} = {\bf C} \, \bv{g},
\end{equation}
where $\bv{T} = [T^{\hat{y}}_{[12]}, T^{\hat{z}}_{[12]}]^\intercal$, $\bv{g} = [\mathcal{G}_1, \mathcal{G}_2]^\intercal/\omega$, and the coefficient matrix
\begin{equation}
    \bf{C} = 
    \begin{pmatrix}
        C_1(\hat{y}, \bv{p}, \bv{q}) & C_2(\hat{y}, \bv{p}, \bv{q}) \\
        C_1(\hat{z}, \bv{p}, \bv{q}) & C_2(\hat{z}, \bv{p}, \bv{q})
    \end{pmatrix}.
\end{equation}

\section{Selected results and discussion} \label{sec:res}
We first present our results for the unpolarised Compton amplitude. Our simulations are carried out on QCDSF/UKQCD-generated $2+1$-flavour gauge configurations. Two ensembles are used with volumes $V=[32^3 \times 64, 48^3 \times 96]$, and couplings $\beta=[5.50, 5.65]$ corresponding to lattice spacings $a=[0.074, 0.068] \, {\rm fm}$, respectively. Quark masses are tuned to the $SU(3)$ symmetric point where the masses of all three quark flavours are set to approximately the physical flavour-singlet mass, $\overline{m} = (2 m_s + m_l)/3$~\cite{Bietenholz:2010jr,Bietenholz:2011qq}, yielding $m_\pi \approx [470, 420] \, {\rm MeV}$. Up to $\mathcal{O}(10^4)$ and $\mathcal{O}(10^3)$ measurements are performed by employing multiple sources on the $32^3 \times 64$ and $48^3 \times 96$ ensembles, respectively.

\begin{figure}[ht]
    \centering
    \includegraphics[width=.49\textwidth]{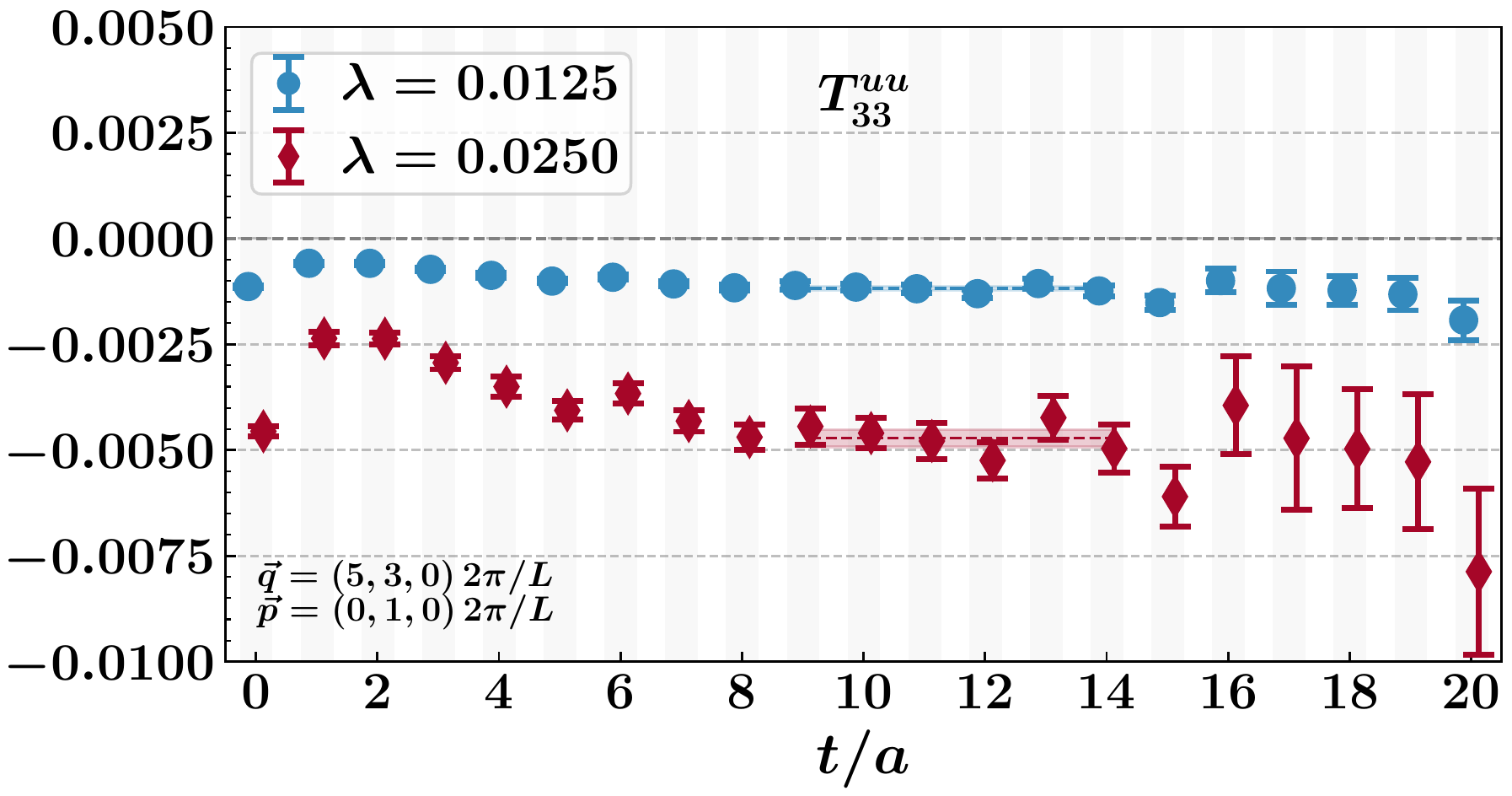}
    \hfill
    \includegraphics[width=.49\textwidth]{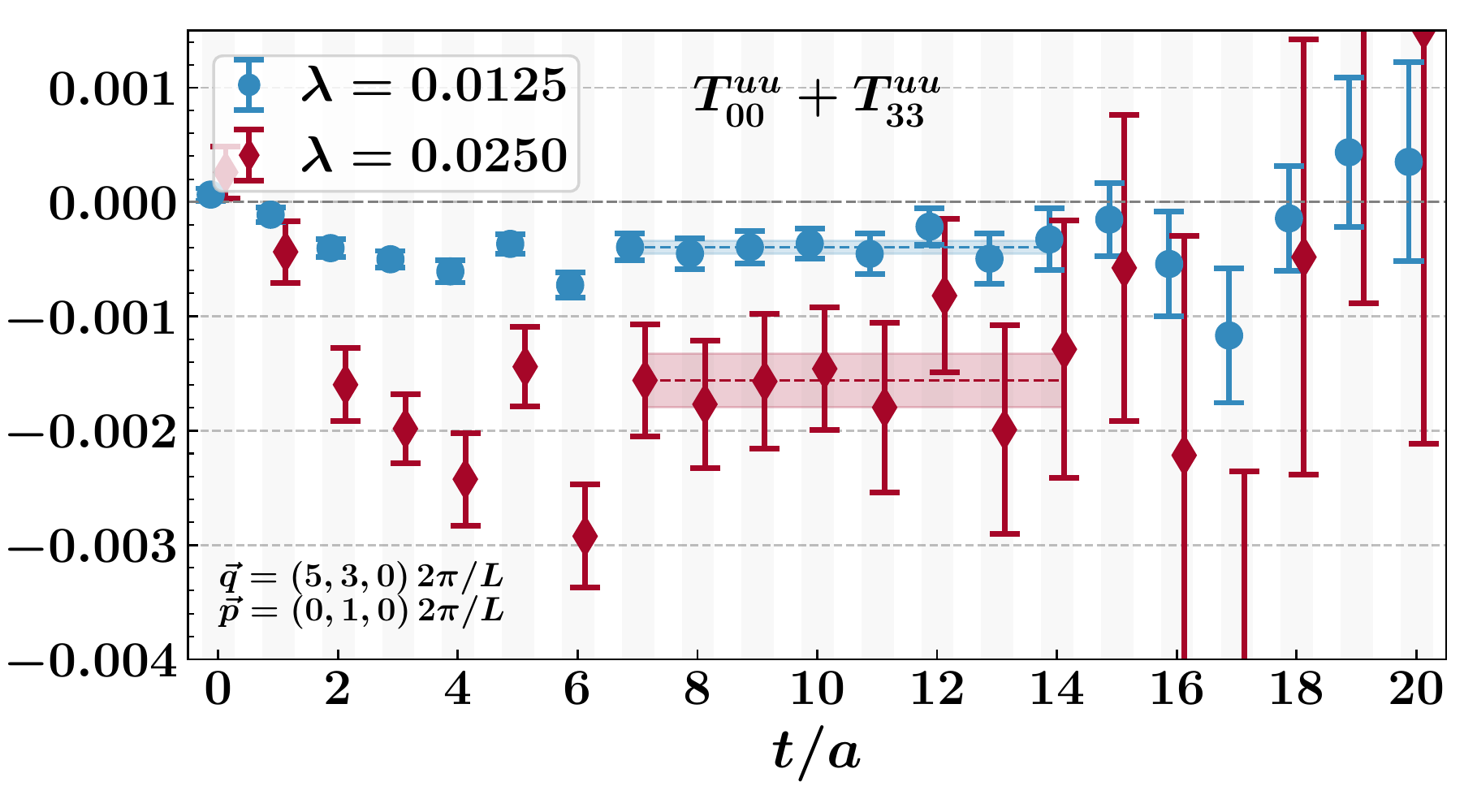}
    \caption{\label{fig:effmass}Effective mass plots of the ratios (\Cref{eq:ratio_fd}) for the amplitudes $T_{33}$ (left) and $T_{00}+T_{33}$ (right). Fit windows, along with the extracted energy shifts with their $1\sigma$ uncertainty, are shown by the shaded bands. We are showing the results obtained on the $48^3 \times 96$ ensemble for the $uu$ piece, for $(\bv{p},\bv{q}) = ((0,1,0),(5,3,0)) \, \latmom$ corresponding to $\omega = 0.18$ at $Q^2 \sim 4.9 \, {\rm GeV}^2$. Figures taken from~\cite{QCDSF:2022btn}. 
    }
\end{figure}

We obtain amplitudes for several values of current momentum, $Q^2$, in the range $1.5 \lesssim Q^2 \lesssim 7$ GeV$^2$. Multiple $\omega$ values are accessed at each simulated value of $\bv{q}$ by varying the nucleon momentum $\bv{p}$, which allows for a mapping of the $\omega$ dependence of the Compton structure functions. For each $\omega$, we extract the energy shifts from the ratios defined in \Cref{eq:ratio_fd,eq:ratio_fm} for two $|\lambda|$ values. We determine the fit windows by a covariance-matrix based $\chi^2$ analysis where we choose the ranges that have $\chi^2_{dof} \sim 1.0$. Effective mass plots for the ratios are shown in \Cref{fig:effmass} for some selected cases.

\begin{figure}[t]
    \centering
    \includegraphics[width=.65\textwidth]{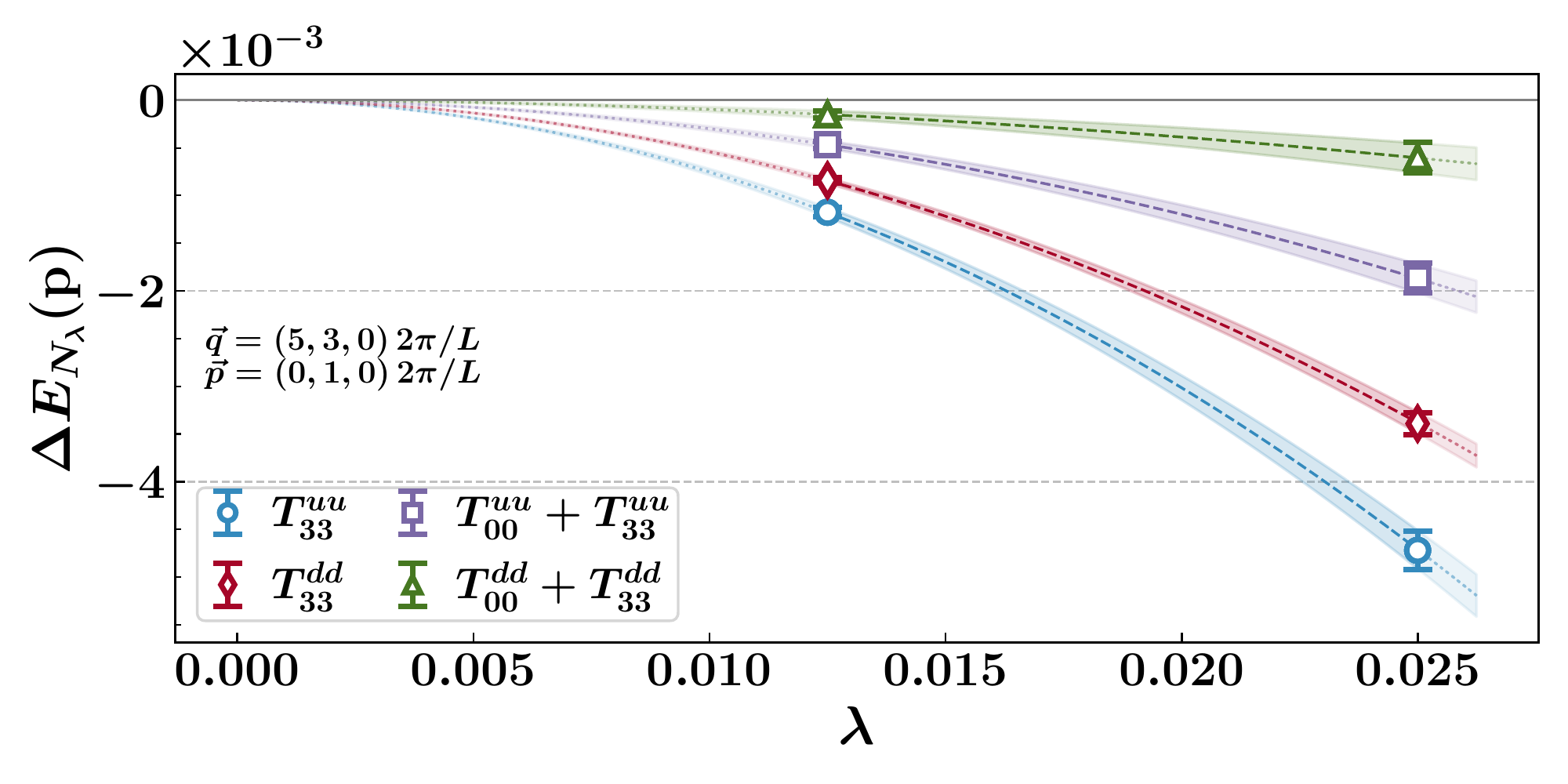}
    \caption{\label{fig:lamfit}$\lambda$ dependence of the energy shifts for the $uu$ and $dd$ pieces of the $T_{33}$ and $T_{00} + T_{33}$ amplitudes. Results are from the $48^3 \times 96$ ensemble with the same kinematics given in \Cref{fig:effmass}. Figure taken from~\cite{QCDSF:2022btn}.
    }
\end{figure}

We perform polynomial fits of the form, $\Delta E_{N_\lambda}(\bv{p}) = \lambda^2 \left. \frac{\partial^2 E_{N_\lambda}(\bv{p})}{\partial \lambda^2} \right|_{\lambda = \bv{0}} + \mathcal{O}(\lambda^4)$, to determine the second order energy shift. \Cref{eq:enshift_oo} also reduces to this form since we set $|\lambda_1| = |\lambda_2| = |\lambda|$ in our simulations. The unperturbed energy, $E_N$, and odd-order lambda terms ($\mathcal{O}(\lambda)$, $\mathcal{O}(\lambda^3)$, $\dots$) are removed by construction in the ratios. Given the smallness of our $\lambda$ values, higher order $\mathcal{O}(\lambda^4)$ terms are heavily suppressed, hence the fit form reduces to a simple one parameter polynomial.  We show representative cases for the $\lambda$ fits in \Cref{fig:lamfit}. We confirm the suppression of the $\mathcal{O}(\lambda^4)$ term, and the absence of $\lambda$-odd terms, by including $\mathcal{O}(\lambda)$, $\mathcal{O}(\lambda^3)$, and $\mathcal{O}(\lambda^4)$ terms separately in the fit. We find that any residual contamination has a negligible effect compared to the statistical error on the extracted amplitudes.

\begin{figure}[ht]
    \centering
    \includegraphics[width=\textwidth]{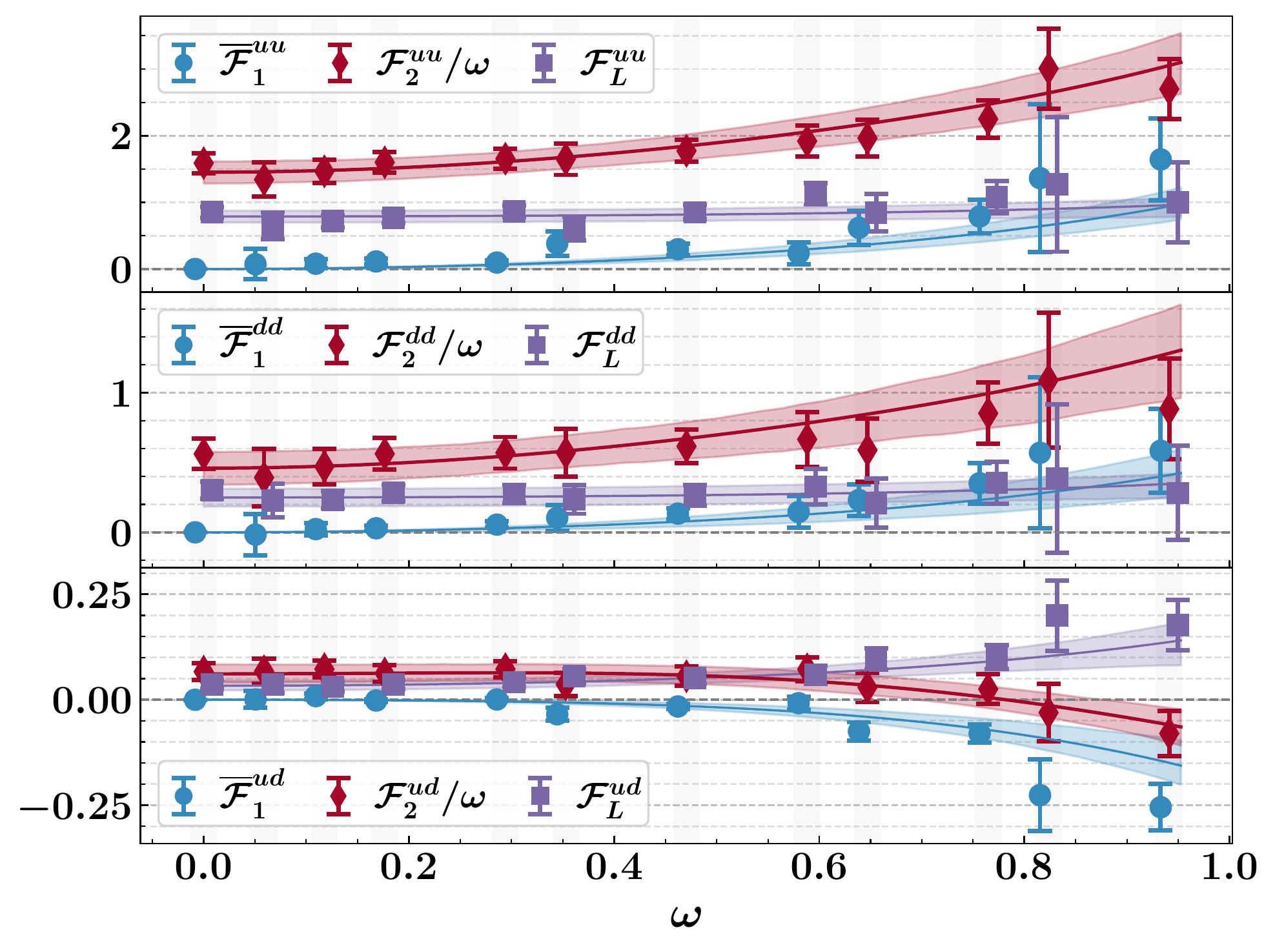}
    \caption{\label{fig:F12L}$\omega$ dependence of the Compton structure functions $\overline{\mathcal{F}}_1$, $\mathcal{F}_2$, and $\overline{\mathcal{F}}_L$ at $Q^2 \sim 4.9 \, {\rm GeV}^2$ for the $uu$ (top), $dd$ (middle) and $ud$ (bottom) contributions. Coloured shaded bands show the fits with their 68\% credible region of the highest posterior density. Points are displaced for clarity. Figure taken from~\cite{QCDSF:2022btn}.
    }
\end{figure}

The above analysis is performed to map out the $\omega$ dependence of the Compton structure functions given in \Cref{eq:compF1,eq:compF2} for each $Q^2$ value that we study. $\mathcal{F}_L(\omega,Q^2)$ is constructed according to \Cref{eq:FL_comp}. We show the $\omega$ dependence of the Compton structure functions, along with their fit curves, in \Cref{fig:F12L} for a representative case of $Q^2 \sim 4.9 \, {\rm GeV}^2$ calculated on the $48^3 \times 96$ ensemble.

The first few Mellin moments of $F_1$, $F_2$ and $F_L$ are determined by performing a simultaneous fit to $\overline{\mathcal{F}}_1$ and $\mathcal{F}_2$ in a Bayesian framework at each $Q^2$ value. We use \Cref{eq:ope_moments1} for $\overline{\mathcal{F}}_1$ and express $\mathcal{F}_2$ in terms of the independently positive definite moments of $F_1$ and $F_L$,
\begin{align}\label{eq:moments_F2}
    \frac{\mathcal{F}_2(\omega)}{\omega} = \frac{\tau}{\left(1 + \tau \, \omega^2 \right)} \sum_{n=0}^{\infty} 4\omega^{2n} \left[ M_{2n}^{(1)} + M_{2n}^{(L)} \right],
\end{align}
where $\tau=Q^2 / 4 M_N^2$, $M^{(1)}_{0}(Q^2) = 0$, and $M^{(L)}_{0}(Q^2) = \frac{4M_N^2}{Q^2} M^{(2)}_{2}(Q^2)$. The intercept at $\omega=0$ is proportional to the lowest moment of $F_2$, i.e. $M_2^{(2)}(Q^2)$. We truncate the series at $n=4$ (inclusive). No dependence on higher-order terms is seen. We sample the moments from uniform distributions with bounds $M_{2}(Q^2) \in [0,1]$ and $M_{2n}(Q^2) \in [0, M_{2n-2}(Q^2)]$, for $n > 1$, to enforce the monotonic decreasing nature of the moments, $M_2(Q^2) \ge M_4(Q^2) \ge \cdot \cdot \cdot \ge M_{2n}(Q^2) \ge \cdot \cdot \cdot \ge 0$, for $uu$ and $dd$ contributions separately. Note that the positivity bound does not hold for the $ud$ contributions but they are constrained by $\left| M_{2n}^{ud}(Q^2) \right|^2 \le 4 M_{2n}^{uu}(Q^2) M_{2n}^{dd}(Q^2)$, since the total inclusive cross section (hence each moment) is positive for any value of the quark charges and at all kinematics. The sequences of individual $uu$, $dd$, and $ud$ moments are selected according to the standard probability distribution, $\operatorname{exp}(-\chi^2/2)$, where,
$\chi^2 = \sum_{\mathcal{F}} \sum_{i} \left[ \mathcal{F}^\text{model}_i - \mathcal{F}^\text{obs}(\omega_i) \right]^2 / \sigma^2$,
is the $\chi^2$ function with $\sigma^2$ the diagonal elements of the full covariance matrix. Here, $\mathcal{F}$ stands for $\overline{\mathcal{F}}_1$ and $\mathcal{F}_2$, and the indices $i$, $j$ run through all the $\omega$ values and flavour-diagonal and mixed-flavour pieces. We account for the correlations between the data points by a bootstrap analysis. Fits depicting the extraction of the moments are also shown in \Cref{fig:F12L} by shaded bands for a representative case.

\begin{figure}[h]
    \centering
    \includegraphics[width=.9\textwidth]{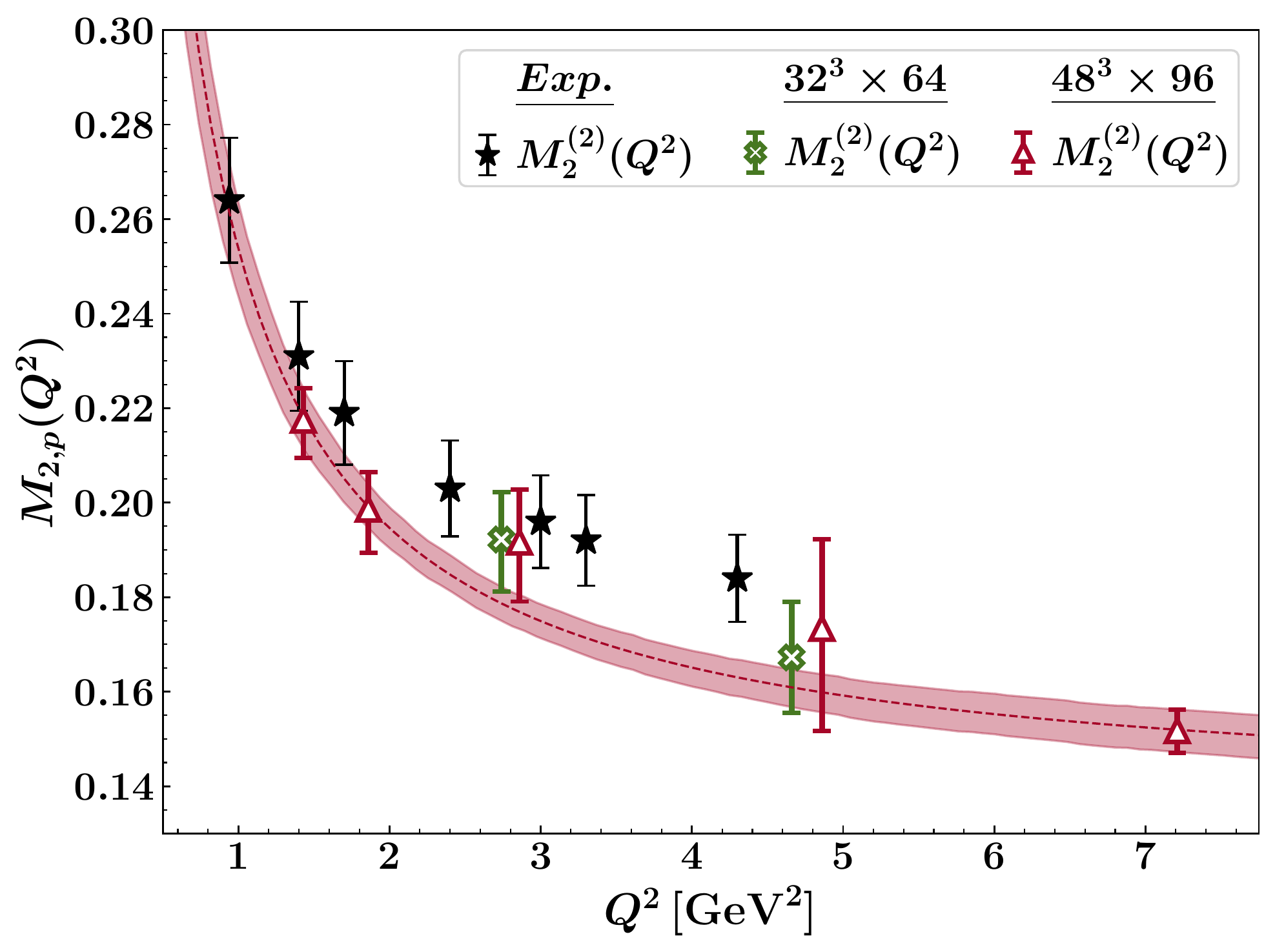}
    \caption{\label{fig:F2_moments}$Q^2$ dependence of the lowest moments of proton $F_{2}$. Filled stars are the experimental Cornwall-Norton moments of $F_2$~\cite{Armstrong:2001xj}. Figure taken from~\cite{QCDSF:2022btn}.
    }
\end{figure}

We show the lowest moments of $F_{2}$ for proton in \Cref{fig:F2_moments} as a function of $Q^2$. Note that the moments of the proton are constructed via $M_{2,p}^{(2,L)} = \frac{4}{9} M_{2,uu}^{(2,L)} + \frac{1}{9} M_{2,dd}^{(2,L)} - \frac{2}{9} M_{2,ud}^{(2,L)}$. Also shown are the experimental determinations of the Cornwall-Norton moments of $F_{2}$~\cite{Armstrong:2001xj}. We see a remarkable agreement, although we should note that our systematics are not fully accounted for yet.   

The Compton amplitude encompasses all power corrections, therefore it is possible to estimate the leading power correction (i.e. twist-4) by studying the $Q^2$ behaviour of the moments in a twist expansion,
\begin{equation}\label{eq:ht_model}
    M_{2,h}^{(2)}(Q^2) = M_{2,h}^{(2)} + C_{2,h}^{(2)}/Q^2 + \mathcal{O}(1/Q^4),    
\end{equation}
where $h \in \{uu, dd, ud, p\}$. Utilising only the $M_2^{(2)}(Q^2)$ moments obtained on the $48^3 \times 96$ ensemble, we study the power corrections down to $Q^2 \approx 1.5 \; {\rm GeV}^2$. Our fit for proton is shown in \Cref{fig:F2_moments}. The extracted values for $M_{2,h}^{(2)}$ and $C_{2,h}^{(2)}$ are collected in \Cref{tab:model}. Although we focus on the proton due to availability of experimental data, it is possible to estimate the moments for neutron or the isovector $u-d$ combination since we have the contributions of different flavour components. Power corrections are a combination of target mass corrections, pure higher-twist terms, and the elastic contributions. These effects can be disentangled further, for instance by determining the elastic contributions from form factors~\cite{Carlson:1998gf,Melnitchouk:2001eh}, and employing Nachtmann moments~\cite{NACHTMANN1973237} to account for the target mass corrections, along with including the logarithmic evolution of moments in \Cref{eq:ht_model}.  We leave such an investigation to a future study. 
\begin{wraptable}{l}{.4\textwidth}
\centering
\caption{\label{tab:model} Asymptotic values of the moments and the coefficients of the leading power correction terms. We quote the power corrections at the scale of the nucleon mass $Q^2 = M_N^2$.}
    \begin{tabular}{lcc}
        \hline\hline
        $h$ & $M_{2,h}^{(2)}$ & $C_{2,h}^{(2)} / M_N^2$ \\
        \hline
        $uu$ & 0.268(13) & 0.206(24) \\ 
        $dd$ & 0.146(7)  & 0.024(14) \\ 
        $ud$ & 0.000(0)  & 0.007(3)  \\ 
        $p$  & 0.135(6)  & 0.091(11) \\ 
        \hline\hline
    \end{tabular}
\end{wraptable} 

\begin{figure}[ht]
    \centering
    \includegraphics[width=.9\textwidth]{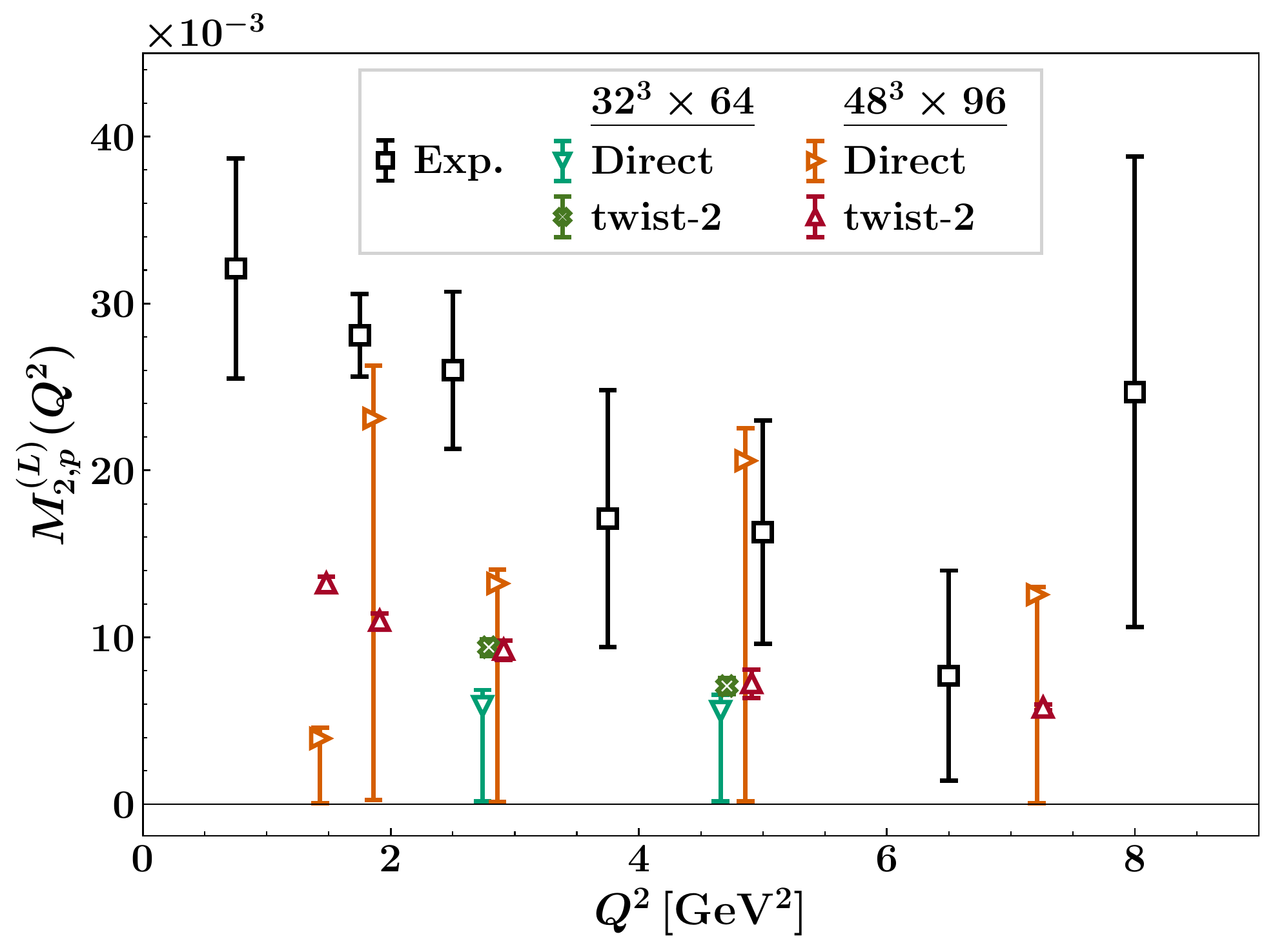}
    \caption{\label{fig:FL_moments}$Q^2$ dependence of the lowest moments of proton $F_{L}$ (Direct). Open squares are the experimental Nachtmann moments of $F_L$~\cite{Monaghan:2012et}. We also show the moments (twist-2) determined via the relation, \Cref{eq:FL_qcd}. Twist-2 points are displaced for clarity. Figure taken from~\cite{QCDSF:2022btn}.
    }
\end{figure}

In \Cref{fig:FL_moments}, we show the lowest (Cornwall-Norton) moments of $F_L$ in comparison to the experimentally determined Nachtmann moments~\cite{Monaghan:2012et}. With our current precision, we are able to set an upper bound for the moments that is compatible with the experimental moments. 

The leading twist part of $F_L$ is related to $F_2$ in leading-order in $\alpha_s$. In terms of moments this relation reads~\cite{ALTARELLI197889},
\begin{equation}\label{eq:FL_qcd}
    M_{2,p}^{(L),{\rm twist-}2}(Q^2) = \frac{4}{9\pi} \alpha_s(Q^2) M_{2,p}^{(2),{\rm twist-}2}(Q^2),
\end{equation}
where we have replaced the leading-twist moment on RHS with $M_{2,p}^{(2)}(Q^2)$ from the current work as an approximation. We use the value of $\alpha_s(Q^2)$ determined in the $\overline{MS}$ scheme at $\mu=Q^2$ at the four-loop order. 
The $Q^2$ behaviour is in good agreement with experimental points as shown in \Cref{fig:FL_moments}. Improving the precision in future studies, would help to resolve the difference between the direct determination and twist-2 part of the lowest few moments of $F_L$ and reveal the higher-twist effects. 
\begin{figure}[ht]
    \centering
    \includegraphics[width=\textwidth]{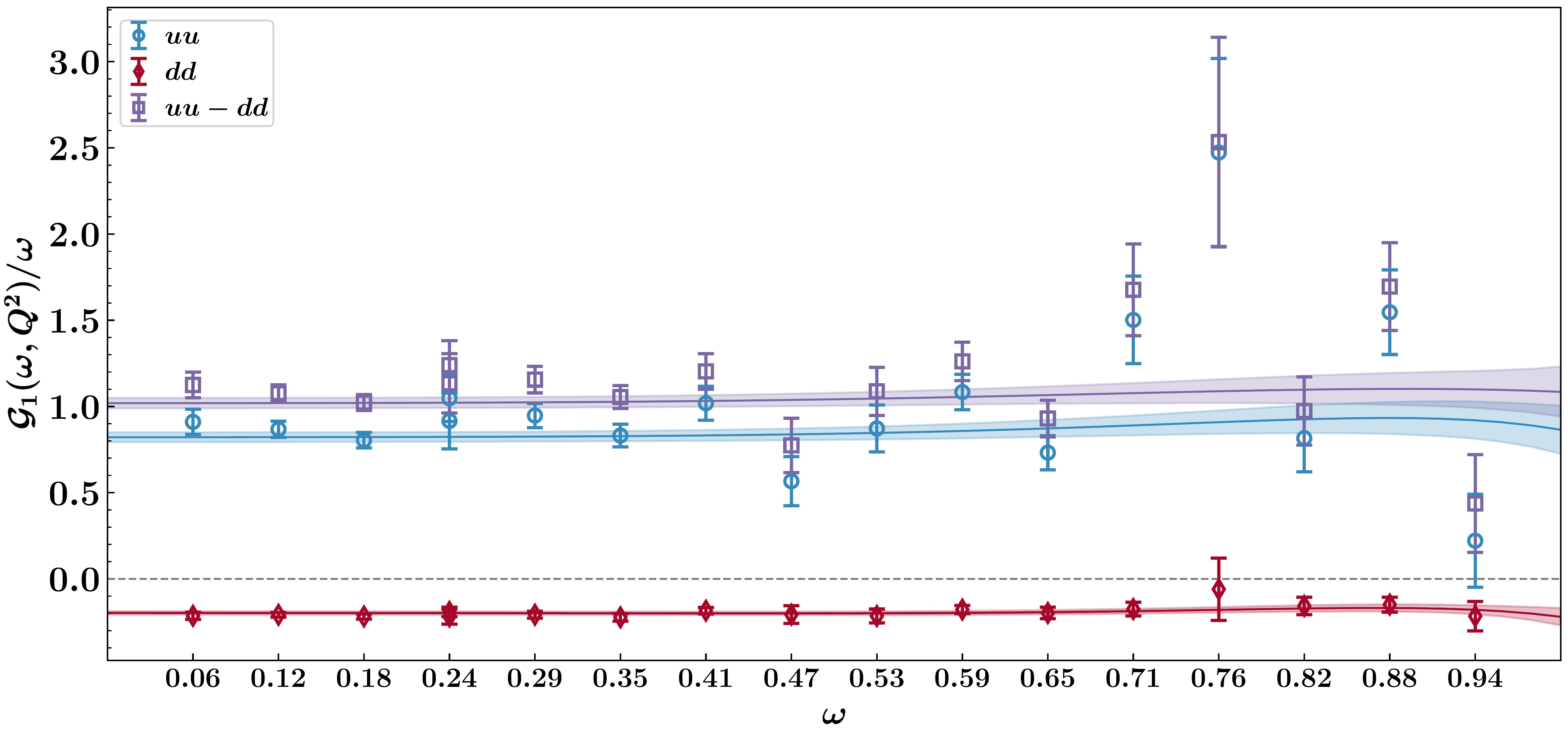}
    \caption{\label{fig:g1} The polarised Compton structure function $\mathcal{G}_1(\omega,Q^2)/\omega$ as a function of $\omega$. We show the preliminary result obtained on the $48^3 \times 96$ ensemble at $Q^2 \sim 4.9 \, {\rm GeV}^2$. 
    }
\end{figure}

Finally, in \Cref{fig:g1} we show the the $uu$ and $dd$ pieces of the polarised $\mathcal{G}_1$ Compton structure function, along with the isovector $uu - dd$ combination, as a function of $\omega$. We extract the first few Mellin moments using \Cref{eq:ope_moments1_pol} by following the same Bayesian analysis performed for the unpolarised case. Shaded bands in \Cref{fig:g1} depict the resulting fits.

The lowest moment of $g_1$ is a particularly interesting phenomenological quantity. It is directly related to the matrix elements of the axial current at leading-twist. For instance, the leading-twist contribution of the lowest isovector moment $\tilde{M}_{2,uu-dd}^{(1)}$ provides a complementary approach to determining the nucleon isovector axial charge $g_A$ via the Bjorken sum rule~\cite{Manohar:1992tz,Larin:1991tj}, or alternatively, the QCD effective charge~\cite{Deur:2016tte,Deur:2022msf}. However, our current results on polarised structure functions are very preliminary. They are obtained on a single ensemble at a single photon virtuality, $Q^2 \sim 4.9 \, {\rm GeV}^2$ with limited statistics in an exploratory simulation. Although these are encouraging results, more progress is needed. 

\section{Summary and outlook}
The Compton amplitude approach has reached a certain maturity where it is possible to directly investigate the structure functions including the effects beyond leading twist. We have overviewed the relations between the forward Compton amplitude and the structure functions, and described a novel extension of the Feynman-Hellmann techniques that simplifies the calculation of the amplitude. We showed the versatility of this approach by calculating the moments of transverse and longitudinal proton structure functions along with their $Q^2$ dependence. This allows us to study the power corrections for the first time in a lattice calculation. 
Currently our calculations involve configurations with two different lattice spacings and volumes, all at the $SU(3)$ symmetric point. Calculations on additional ensembles that cover a range of lattice spacings and pion masses are required to fully account for systematic effects.

We are working towards extending our formalism to include the spin-dependent structure functions. Our preliminary results are encouraging. Additionally, accessing the parity violating structure function $F_3$, by considering weak currents is an exciting future direction.

\acknowledgments
I thank the organisers of ``The 39th International Symposium on Lattice Field Theory'' for the invitation. I would also like to thank my colleagues from the QCDSF/UKQCD collaboration, this work would not have been possible without their contributions. The numerical configuration generation (using the BQCD lattice QCD program~\cite{Haar:2017ubh})) and data analysis (using the Chroma software library~\cite{Edwards:2004sx}) was carried out on the DiRAC Blue Gene Q and Extreme Scaling (EPCC, Edinburgh, UK) and Data Intensive (Cambridge, UK) services, the GCS supercomputers JUQUEEN and JUWELS (NIC, Jülich, Germany) and resources provided by HLRN (The North-German Supercomputer Alliance), the NCI National Facility in Canberra, Australia (supported by the Australian Commonwealth Government) and the Phoenix HPC service (University of Adelaide). KUC is supported by the Australian Research Council grants DP190100297 and DP220103098. For the purpose of open access, the authors have applied a Creative Commons Attribution (CC BY) licence to any Author Accepted Manuscript version arising from this submission.

\bibliographystyle{JHEP}
\begingroup
    \renewcommand{\baselinestretch}{1}
    \setlength{\bibsep}{5pt}
    \setstretch{1}
    \bibliography{proceeding}
\endgroup

\end{document}